\documentclass[aps,prl,floatfix,twocolumn,footinbib,superscriptaddress]{revtex4-1}
\usepackage{epsfig}
\usepackage{graphicx}
\usepackage{dcolumn}   
\usepackage{bm}
\usepackage{amssymb}
\usepackage{mathrsfs}
\usepackage{amsmath}
\usepackage{bbold}
\usepackage{epstopdf}

\begin{document}

\title{Angular momentum radiation from current-carrying molecular junctions}

\preprint{APS/123-QED}
\author{Zu-Quan Zhang}
\email{phyzhaz@nus.edu.sg}
\affiliation{Department of Physics, National University of Singapore, Singapore 117551, Republic of Singapore}
\author{Jing-Tao L\"u}
\email{jtlu@hust.edu.cn}
\affiliation{School of Physics and Wuhan National High Magnetic Field Center, Huazhong University of Science and Technology, 430074 Wuhan, P. R. China}
\author{Jian-Sheng Wang}
\email{phywjs@nus.edu.sg}
\affiliation{Department of Physics, National University of Singapore, Singapore 117551, Republic of Singapore}

\date{09 January 2019}


\begin{abstract}
We consider the radiation of angular momentum (AM) from current-carrying molecular junctions. Using the nonequilibrium Green's function method, we derive a convenient formula for the AM radiation and apply it to a prototypical benzene molecule junction. We discuss the selection rules for inelastic transitions between the molecular angular momentum eigenstates due to a 6-fold rotational symmetry. Our study provides important insights into the generation of light with AM from DC-biased molecular junctions. 
\end{abstract}

\maketitle

{\emph{Introduction.--}}
The atomic scale interaction of nonequilibrium electrons with light is the key to develop electrically driven single molecular light sources for sensing, spectroscopy and chemical reactions~\cite{galperin2012,aradhya2013,Kuhnke2017}. 
The high spatial resolution and local field enhancement offered by the tip of a scanning tunneling microscope provide an ideal platform to investigate this problem~\cite{Berndt1991,Berndt1993,Aizpurua2002,qiu2003vibrationally,Dong2004,Schneider2010}. Recent years have witnessed tremendous progress in this direction. By analyzing the light emission spectra, a variety of physical and chemical information can be deduced, including vibrational coupling~\cite{qiu2003vibrationally,Chen2010}, coherent inter-molecular dipole interaction~\cite{Dong2004,zhang2016,imada2016}, plasmon-exciton coupling~\cite{zhangsubnanometre2017,Imada2017}, anti- and super-bunching  photon statistics~\cite{zhang_electrically_2017,leon_photon_2019}, charge and spin state emission~\cite{doppagne_electrofluorochromism_2018,Miwa2019}. Theoretical understanding of these effects relies on methods developed in quantum optics and quantum transport~\cite{Galperin2005,Schneider2012,Lulight2013,Xu2014,Kaasbjerg2015,Nian2018,kroger2018,Miwa2019,Parzefall_2019,mukamel2019flux}.

The coupling of electron orbital motion with its spin leads to spin-orbit interaction, which is of vital importance in spintronics~\cite{Awschalom09}, topological physics~\cite{Hasan2010,Qi2011} and so on. Spin-orbit coupling is also responsible for the chiral induced spin selectivity in electron transport through molecules~\cite{ray_asymmetric_1999,dalum_theory_2019}. However, the effect of electron orbital motion on single molecule electro-luminescence is, to a large extent, unexplored. In this work, based on nonequilibrium Green's function (NEGF) method~\cite{HaugAndJauho,bruus2004many,Wang2008,Wang2014}, we develop a microscopic theory to study electrically driven angular momentum emission from a single molecule. We consider a prototypical benzene molecule, which has well-defined orbital angular momentum states in isolated situation. We illustrate the underlying mechanism as inelastic electronic transition between states with different orbital AM. 
This is in contrast to the optical approach by passing normal light through constructed optical structures~\cite{Zhang2011,Gorodetski2013}. 

{\emph{ Theory.--}}
To consider an open system for light emission, we decompose it into four parts: a molecular system as a quantum emitter, the coupling of the quantum emitter with the radiation field, the radiation field itself, the leads and their couplings with the quantum emitter for pumping energy and electrons into the quantum emitter. 
\begin{figure}
\centering
\includegraphics[width=6 cm]{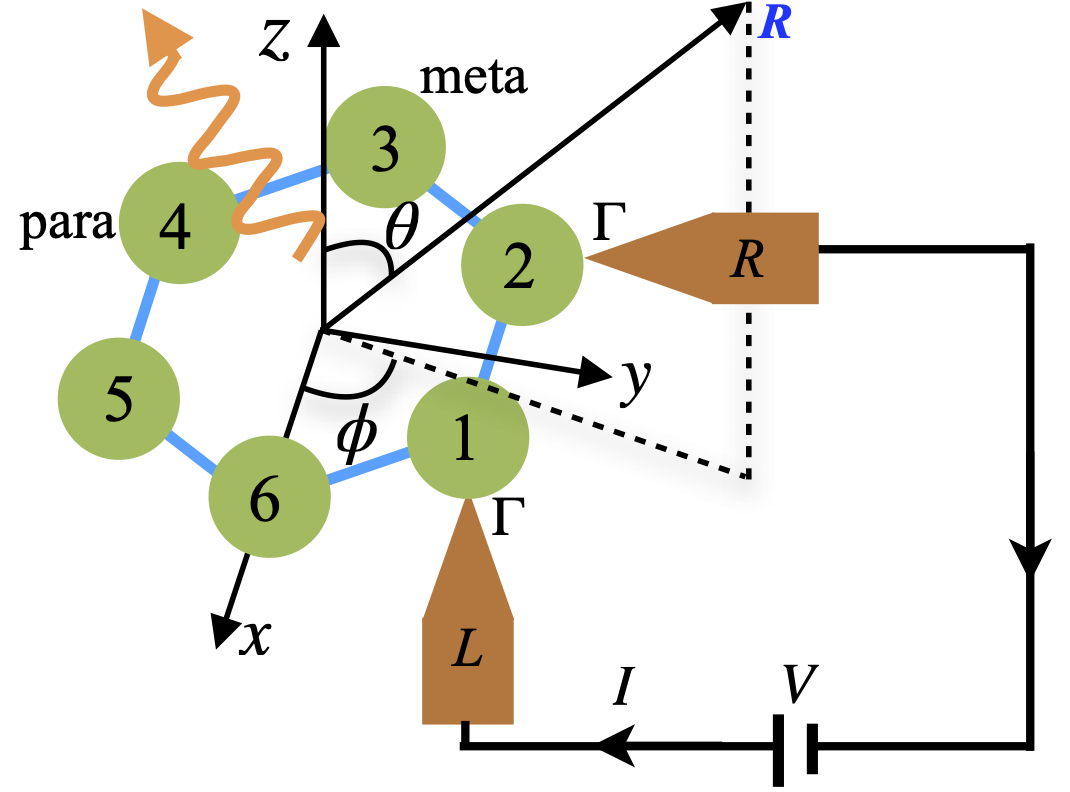}
\caption{Light emission from a benzene molecule junction. The metal leads $L$ and $R$ are connected to two carbon atoms in the ortho position. Meta and para positions correspond to lead $R$ connecting to 3 and 4, respectively. 
}
\label{fig:MD}
\end{figure}

We use a tight-binding (TB) model combined with the Peierls substitution~\cite{Graf1995} to describe the central molecule and its coupling with the radiation field, written as  
\begin{equation} \label{eq:Ht}
H_{t}  =  \sum_{\langle ij \rangle} c_{i}^{\dag} t_{ij} c_{j} e^{i \theta_{ij} },
\end{equation}
where $\langle i j \rangle$ denotes the nearest-neighbor (NN) sites $i$ and $j$, $t_{ij}$ is the NN hopping parameter, $c_{i}^{\dag}$ $(c_{i})$ is the electron creation (annihilation) operator at site $i$. The phase $\theta_{ij} = \frac{e}{\hbar} \int_{\bm{r}_{j}}^{\bm{r}_{i}} \bm{A} \bm{\cdot} d \bm{l}$ represents the coupling to the radiation field. Here,  $e \approx -1.602 \times 10^{-19} \,\textrm{C}$ is the electron's charge, $\bm{A}$ is the vector potential, $\bm{r}_{i}$ and $\bm{r}_{j}$ are the positions of sites $i$ and $j$ respectively. Expanding $e^{i \theta_{ij}}$ in terms of $\bm{A}$ up to first order, we can write Eq.~(\ref{eq:Ht}) into two terms $H_{t} \approx H_{t}^{0} + H_{\textrm{int}}$, with $H_{t}^{0}$ for the noninteracting electrons and $H_{\textrm{int}}$ for the coupling of the  electrons with the radiation field. They are given by
\begin{eqnarray} 
H_{t}^{0} =&&  \sum_{\langle i j \rangle} c^{\dag}_{i} t_{ij} c_{j}, \label{eq:Ht0} \\
H_{\textrm{int}} \approx&& \sum_{\langle ij \rangle} \sum_{k} \sum_{\mu = x, y, z} M_{ij}^{k\mu} c_{i}^{\dag}  c_{j} A_{\mu}(\bm{r}_{k}), \label{eq:HJA}
\end{eqnarray}
where $M_{ij}^{k \mu} = i \frac{e}{2 \hbar} t_{ij} (\bm{r}_{i} - \bm{r}_{j})_{\mu} (\delta_{ki} + \delta_{kj})$ is the electron-photon coupling matrix element. We use Greek letters to represent components of the Cartesian coordinates, i.e.,  $\mu = x, y, z$.

Hamiltonian of the radiation field is 
\begin{equation} \label{eq:Hrad}
H_{\textrm{rad}} = \frac{1}{2} \int d^3 \bm{r} \big( \varepsilon_{0} \bm{E}_{\bot}^2 +  \frac{1}{\mu_{0}} \bm{B}^2 \big),
\end{equation}
where $\varepsilon_{0}$ and $\mu_{0}$ are the vacuum permittivity and permeability, respectively. Here, we adopt the Coulomb gauge with $\nabla\cdot \bm{A}=0$, thus the transverse electric field is given by $\bm{E}_{\bot} = - \partial_{t} \bm{A}$, and the magnetic field is $\bm{B} = \bm{\nabla} \times \bm{A}$. We restrict our discussion to the far-field radiation here and neglect the longitudinal electric field, which is important only in the near-field heat transfer~\cite{Volokitin2007, ZhangEnergy2018, WangCoulomb2018}. The effect of the leads and their couplings with the molecule are included by the self-energies, as shown below. 

The energy and AM flux of electromagnetic field can be obtained from~\cite{Barnett2002, JanowiczQuantum2003}
\begin{eqnarray} 
\bm{S} &=& \frac{1}{\mu_{0}} \big\langle : \bm{E}_{\bot} \times \bm{B}  \colon \big\rangle,  \label{eq:Poynting}  \\
\overleftrightarrow{\bm{\mathcal{M}}} &=& \langle \colon \bm{r} \times \overleftrightarrow{{\bm{T}}} \colon \rangle, \label{eq:AMF}
\end{eqnarray}
where $\langle \colon AB \colon \rangle$ denotes normal order of operators $AB$ when taking the ensemble average, which removes the zero-point motion contribution, and $\overleftrightarrow{{\bm{T}}}$ is the Maxwell stress tensor with $T_{\mu \nu} = \frac{1}{2} \delta_{\mu \nu} (\varepsilon_{0} E^2 + \mu_{0}^{-1} B^2) - \varepsilon_0 E_{\mu} E_{\nu} - \mu_{0}^{-1} B_{\mu} B_{\nu}$.  Equations~(\ref{eq:Poynting}) and (\ref{eq:AMF}) can be expressed in terms of photon Green's function (GF)~\cite{Supplementdetial}
\begin{equation} \label{eq:Sperp}
\begin{split}
S^{\mu} (\bm{r})  =&\epsilon_{\mu \nu \gamma} \epsilon_{\gamma \delta \xi}  \frac{2}{\mu_{0}}  \int_{0}^{+\infty} \frac{d \omega}{2 \pi} \hbar \omega     \\
& \times \textrm{Re}\bigg[- \frac{\partial}{\partial x_{\delta}' } D_{\nu \xi}^{<}(\bm{r},\bm{r}' , \omega)  \bigg] \bigg{|}_{\bm{r}' \to \bm{r}}. 
\end{split}
\end{equation}
Einstein summation rule is used here, and $\epsilon_{\mu \nu \gamma}$ is the Levi-Civita symbol, $D_{\nu \xi}^{<}(\bm{r},\bm{r}' , \omega) $ is the photon's lesser GF in the frequency domain. Relevant quantities in Eq.~(\ref{eq:AMF}) are written as
\begin{subequations}
\begin{align}  
 \langle \colon E_{\mu} E_{\nu} \colon \rangle = & \textrm{Re} \bigg[ \frac{2 i}{\hbar} \int_{0}^{\infty} \frac{d\omega}{2 \pi} (\hbar \omega)^2    D_{\mu \nu}^{<}(\bm{r},  \bm{r} , \omega) \bigg],  \label{eq:EEGF}   \\
 \begin{split}
\langle \colon B_{\mu} B_{\nu} \colon \rangle = & \textrm{Re} \bigg[ i  2 \hbar \int_{0}^{\infty} \frac{d \omega}{2 \pi} \epsilon_{\mu \gamma \xi}  \epsilon_{\nu \gamma' \xi'} \\
&\times  \frac{\partial}{\partial x_{\gamma}}  \frac{\partial}{\partial x_{\gamma'}' }  D_{\xi \xi'}^{<}(\bm{r},  \bm{r}' , \omega) \bigg] \bigg{|}_{\bm{r}' \to \bm{r}}.   \label{eq:BBGF}
\end{split}
\end{align}
\end{subequations}

The GFs are obtained following the standard NEGF formalism. The retarded GF is solved by the Dyson equation $D^{r} = d^{r} + d^{r} \Pi^{r} D^{r}$, and $d^{r}$ is the free space photon GF~\cite{keller2012quantum}. The lesser GFs  are obtained by the Keldysh equation $D^{<} = D^{r} \Pi^{<} D^{a}$, with $D^{a} = (D^{r})^{\dag}$. We use the random phase approximation to calculate the interacting self-energy 
\begin{widetext}
\begin{equation}  \label{eq:selfint}
 \Pi_{\mu\nu}^{<}(\bm{r}_{i},\bm{r}_{j}, \omega) =  - i \hbar   \int_{-\infty}^{+\infty} \frac{dE}{2\pi \hbar} \textrm{Tr}\Big[ M^{i \mu} g^{<} (E) M^{j \nu} g^{>} (E - \hbar \omega) \Big],  
\end{equation}
\end{widetext}
where $\textrm{Tr}[\cdots]$ means trace over the electron degrees of freedom, $g^{<(>)} = g^{r} \Sigma_{\textrm{leads}}^{<(>)} g^{a}$ is the lesser (greater) GFs for the noninteracting electrons. Here, $\Sigma_{\textrm{leads}}^{<(>)}$ is the lesser (greater) self-energy due to electron's coupling with the leads. The leads are in their respective equilibrium states and the self-energies follow the fluctuation-dissipation theorem, for lead $\beta$, $\Sigma_{\beta}^{<} = i f_{\beta} \Gamma_{\beta}$ and $\Sigma_{\beta}^{>} = i (-1 + f_{\beta}) \Gamma_{\beta}$. Here, $f_{\beta}(E, \mu_{\beta}) = 1 \Big/ \Big[\textrm{exp}\big(\frac{E - \mu_{\beta}}{k_{B} T_{\beta}}\big) + 1\Big]$ is the Fermi distribution function, $\mu_{\beta}$ is the chemical potential, $T_\beta$ is the temperature, $k_B$ is the Boltzmann constant, and $\Gamma_{\beta} = - 2 \textrm{Im} \big[ \Sigma_{\beta}^{r} \big]$ is the coupling-weighted spectrum of the lead. 

To calculate the total energy and AM radiation, we choose a large spherical surface enclosing the molecule, and perform the surface integral
\begin{eqnarray}
P &= & \oint \bm{S} \bm{\cdot} d \bm{A}, \label{eq:RadPower} \\
\frac{d \bm{L}}{d t} &= & \oint \overleftrightarrow{\bm{\mathcal{M}}} \bm{\cdot} d \bm{A},  \label{eq:AMFPower}
\end{eqnarray}
where $d \bm{A} = \hat{\bm{R}} dA$, with $\hat{\bm{R}}= \bm{R}/{R}$ denoting the unit normal vector of the spherical surface $dA$ with radius $R$. In the far-field region (with $R$ much larger than the photon wavelength $\lambda$ and the central molecule's size $\sim$ $a$), we get simplified expressions for Eq.~(\ref{eq:RadPower}) and Eq.~(\ref{eq:AMFPower}) as 
\begin{eqnarray}
P &= & - \int_{0}^{\infty} \frac{d \omega}{2 \pi}  \frac{\hbar \omega^2}{3 \pi \varepsilon_{0} c^3}  \textrm{Im} \big[ \Pi_{\mu\mu} ^{\textrm{tot},<}(\omega) \big], \label{eq:MonoP}  \\
\frac{d L_{\gamma}}{d t} &= & \int_{0}^{\infty}  \frac{d \omega}{2 \pi}   \frac{\hbar \omega}{3 \pi \varepsilon_{0} c^3}  \epsilon_{\gamma \mu \nu } \textrm{Re}\big[ \Pi_{\mu\nu} ^{\textrm{tot},<}(\omega) \big],  \label{eq:MonoDLDt}   
\end{eqnarray}
with $ \Pi_{\mu\nu} ^{\textrm{tot},<}(\omega) = \sum_{i j}  \Pi_{\mu \nu}^{<}(\bm{r}_{i},\bm{r}_{j}, \omega)$. Einstein summation rule is used here. We have performed the solid angle integration within the monopole approximation, i.e., neglecting the molecular size considering $(a / R) \ll 1$. We observe that the emission power is related to the trace, while AM emission rate to the antisymmetric part of the tensor, $\Pi^{\textrm{tot},<}$.  Eq.~(\ref{eq:MonoDLDt}) is the main result of this paper. It serves as a simple formula for calculating far-field radiation of optical AM from biased molecular systems. 

{\emph{ Application.--}}
The theory we develop is quite general. We now apply it to a prototypical benzene molecule junction shown in Fig.~\ref{fig:MD}. We take the NN hopping parameter as $- t_{ij} = t = 2.5 \, \textrm{eV}$ and C-C bond length as $a = 1.4 \, $ \AA.  We use the wideband approximation for molecule-lead coupling with $\Gamma_{L} = \Gamma_{R} = \Gamma$.  

\begin{figure}
\centering
\includegraphics[width=8 cm]{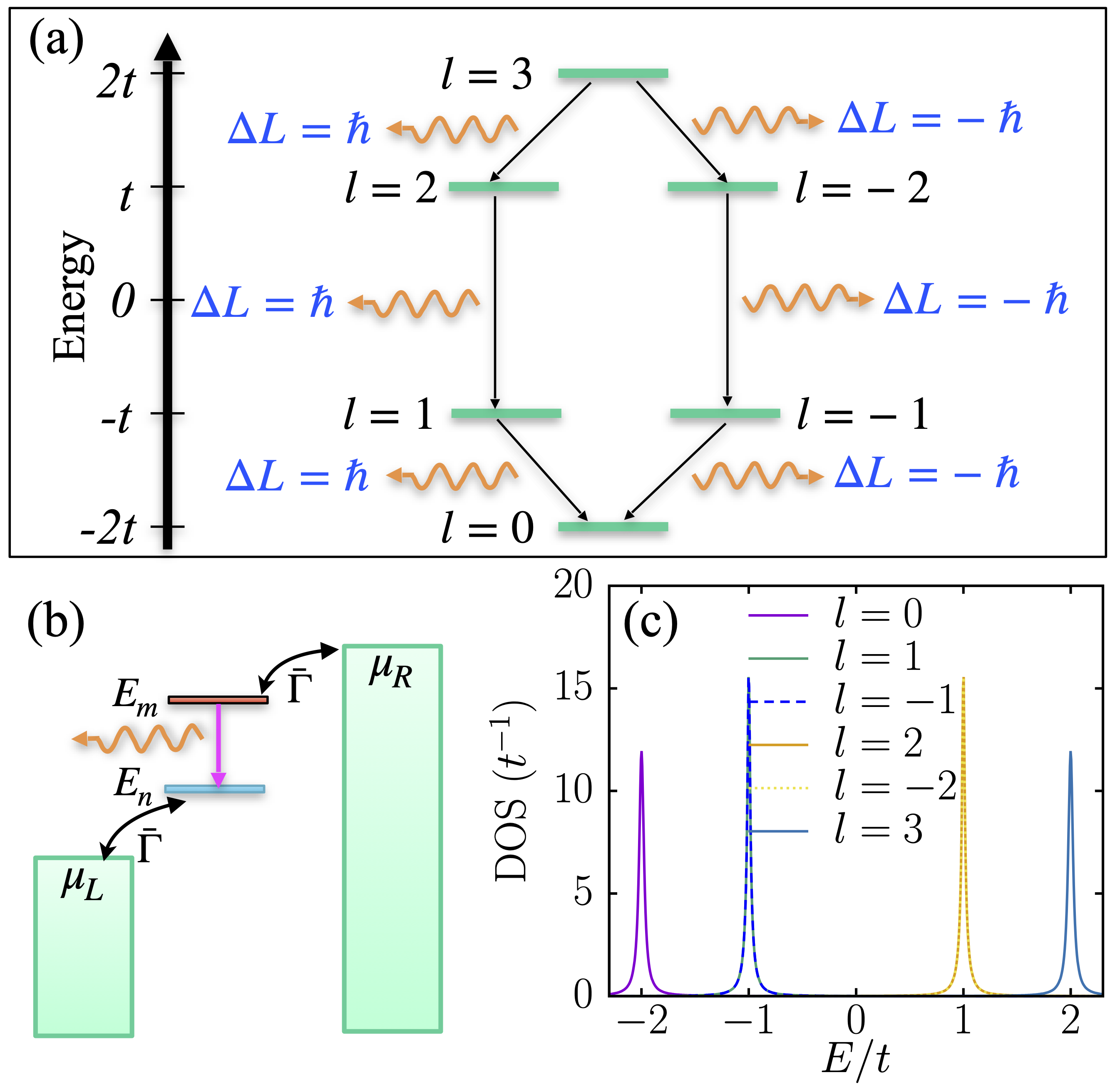}
\caption{(a) Selection rules for the emission of optical angular momentum between energy levels of the TB benzene molecule. (b) Light emission with leads coupled directly to two eigenmodes of the benzene molecule in high bias regime. $\bar{\Gamma}$ is the weak lead coupling to the modes. (c) Orbital resolved electron density of states (DOS) of the benzene molecule when coupled to the leads in ortho position shown in Fig.~\ref{fig:MD}, with $\Gamma = 0.4 \, \textrm{eV}$. The DOS is defined by $- \textrm{Im}\big[\widetilde{g}_{ll}^{r}(E)\big] / \pi$ for energy level $l$. 
}
\label{fig:SRDOS}
\end{figure}

The isolated benzene molecule has a $C_{6}$ rotational symmetry.
We put the molecule in the $x$-$y$ plane with site positions $x_{j} = a \, \textrm{cos} (2\pi j /6)$, $y_{j} = a \, \textrm{sin} (2\pi j /6)$, $j = 1, 2, \dots, 6$.  The orbital eigen energies are $E_{l} = -2 t \, \textrm{cos} (2\pi l / 6)$, with $l = 0, \pm 1, \pm 2, 3$. We have neglected the spin degeneracy here. When coupled to the two metal leads, the energy levels are broadened, but the degeneracy is not lifted. The orbital resolved density of states (DOS) is shown in Fig.~\ref{fig:SRDOS}(c). Here the mode space GF $\widetilde{g}^r$ is related to the real space GF via the unitary transformation $\widetilde{g}^{r} = U^{\dag} g^{r} U$, with $U_{jm} = e^{i 2 \pi j m / 6}/\sqrt{6}$, $j, m = 1, 2, \dots, 6$.  Note that the mode index $m$ is unique only modulo 6, thus $6$ is the same as $0$, and $5$ is the same as $-1$. Also, we use the notation that an operator denoted as $O$ in real space is written as $\widetilde{O}$ in mode space, with $\widetilde{O} = U^{\dag} O U$.

The mechanism of AM emission can be understood by considering
the selection rules in mode space. Defining the electron velocity matrix $v_{ij}^{\mu} =  t_{ij} (\bm{r}_{i} - \bm{r}_{j})_{\mu} / \hbar$, we have $v^{\mu}=\frac{1}{i e} \sum_{k} M^{k \mu}$, which has the $C_6$ rotational symmetry. For the coordinate system we choose here, the $C_6$ symmetry leads to the relations $\widetilde{v}^{x}_{nm} \widetilde{v}^{x}_{mn}=  \widetilde{v}^{y}_{nm} \widetilde{v}^{y}_{mn}$ and $\widetilde{v}^{x}_{nm} \widetilde{v}^{y}_{mn}= i \Delta_{mn} \widetilde{v}^{x}_{nm} \widetilde{v}^{x}_{mn}$. Here, $\Delta_{mn} = \textrm{sgn}(m - n)$ if $|m - n| = 1$, and $\Delta_{16} = - \Delta_{61} = 1$, otherwise $\Delta_{mn} = 0$. 

We consider the simple case where the leads couple respectively to only two eigenmodes of the molecule in the high bias regime $\mu_{L} \ll E_{n} < E_{m} \ll \mu_{R}$ [Fig.~\ref{fig:SRDOS}(b)]. Due to the high bias setting, we have $\widetilde{g}_{mm}^{<}(E) \approx \frac{i \bar{\Gamma}}{(E - E_m)^2 + (\bar{\Gamma}/2)^2}$, $\widetilde{g}_{nn}^{>}(E) \approx \frac{- i \bar{\Gamma}}{(E - E_{n})^2 + (\bar{\Gamma}/2)^2}$, $\widetilde{g}_{mm}^{>}(E) \approx 0$ and $\widetilde{g}_{nn}^{<}(E) \approx 0$. In the limit $\bar{\Gamma} \to 0$, we get from Eq.~(\ref{eq:MonoP}) $P = - \omega_{mn}^2 e^2  ( \widetilde{v}_{nm}^{x} \widetilde{v}_{mn}^{x} + \widetilde{v}_{nm}^{y} \widetilde{v}_{mn}^{y}) / (3 \pi \varepsilon_{0} c^3)$, and $dL_z / d t =  i  \omega_{mn} e^2  (\widetilde{v}_{nm}^{x} \widetilde{v}_{mn}^{y} - \widetilde{v}_{nm}^{y} \widetilde{v}_{mn}^{x}) / (3 \pi \varepsilon_{0} c^3)$ from Eq.~(\ref{eq:MonoDLDt}), with $\hbar \omega_{mn} = E_{m} - E_{n}$~\cite{Supplementdetial}. Using the relations of the velocity matrix due to the $C_6$ symmetry, we get $\frac{dL_z / d t}{P} = \frac{\Delta_{mn}}{\omega_{mn}}$. This result is reminiscent of the Eq.~(19) of a recent work for the classical case of AM radiation from a single electron performing circular motion with a constant frequency~\cite{Katoh2017}. Since every emitted photon carries an energy $\hbar \omega_{mn}$, the number of photons emitted per unit time is $d N / d t = P / (\hbar \omega_{mn})$. Thus, the AM per emitted photon is $\Delta L = \frac{dL_z / dt}{dN / d t} = \Delta_{mn} \hbar$, which is the selection rules shown in Fig.~\ref{fig:SRDOS}(a).  

The light emission for real space coupling in Fig.~\ref{fig:MD} is a combination of all the possible processes shown in Fig.~\ref{fig:SRDOS}(a). This is tuned by the applied bias. Significant light emission between two energy levels of the molecule is possible when the energy levels enter into the bias window, restricted by the selection rules. Since the degenerate energy levels ($l = \pm 1, \pm 2$) are broadened but not split when the molecule couples to the leads, they will enter into or out of the bias window simultaneously. Light emission from inelastic transition $l = 2 \to l = 1$ is accompanied by emission from transition $l = -2 \to l = -1$. These two processes emit light with opposite AM. The summation of the two leads to  cancellation of the total AM. 

\begin{figure}
\centering
\includegraphics[width=8.5 cm]{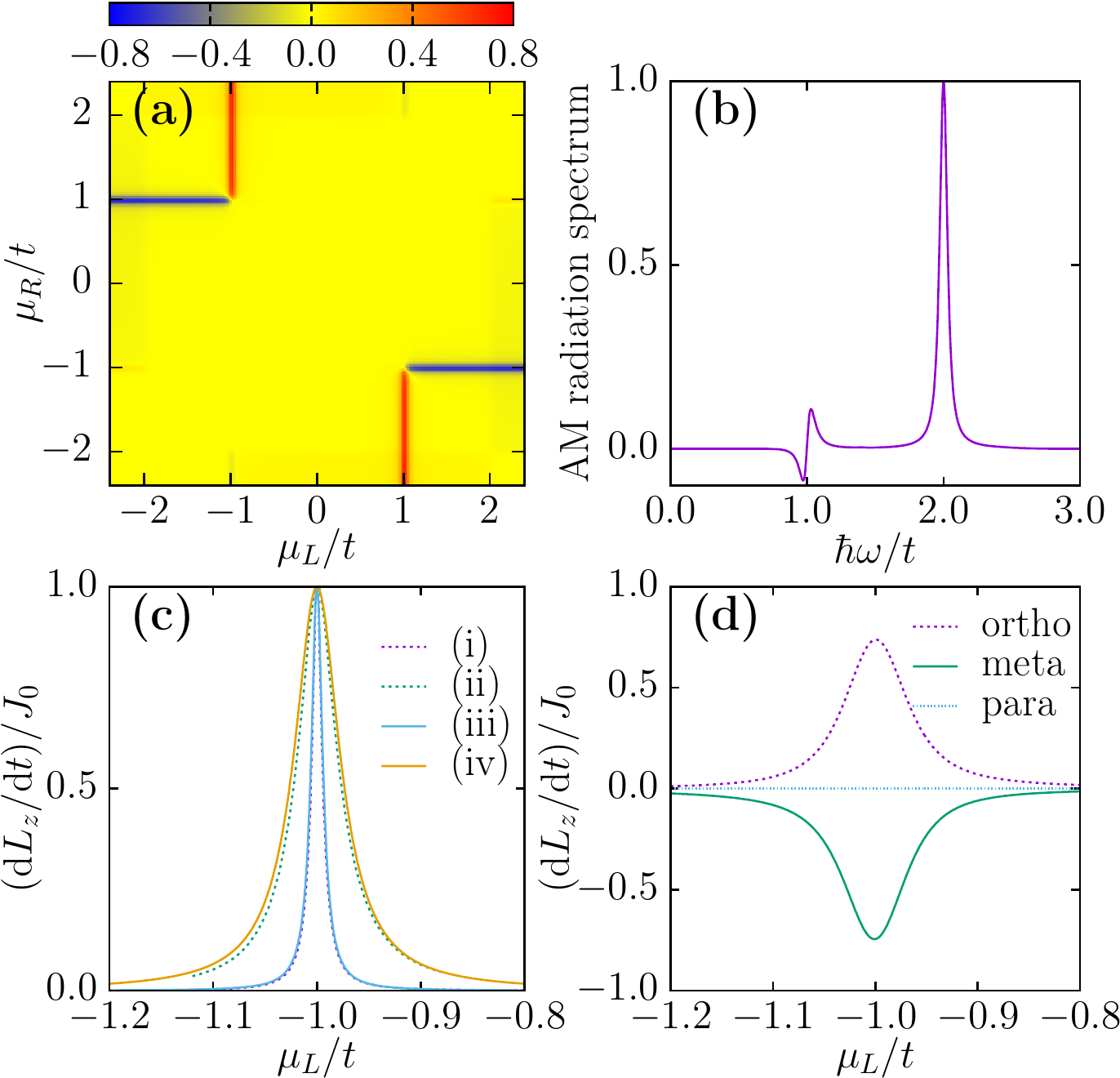}
\caption{(a) Intensity plot of angular momentum radiation rate normalized by $(dL_{z} / dt) / J_{0}$ as a function of chemical potentials $\mu_{L}/t$ and $\mu_{R}/t$. (b) Frequency resolved spectrum of the angular momentum radiation, normalized relative to the maximum value at $\hbar \omega = 2 t$, for $\mu_{L} = -t$ and $\mu_R = 6 \, \textrm{eV}$. (c) Line cut of the plot in (a) at $\mu_{R} = 4 \, \textrm{eV}$. (\romannumeral1)(\romannumeral2) are results calculated from Eq.~(\ref{eq:MonoDLDt}) at zero temperature, while (\romannumeral3)(\romannumeral4) are results from Eq.~(\ref{eq:RET0}). $\Gamma = 0.1 \, \textrm{eV}$ for (\romannumeral1)(\romannumeral3), and $\Gamma = 0.4 \, \textrm{eV}$ for (\romannumeral2)(\romannumeral4). The leads couple to the benzene molecule in ortho position for (a)(b)(c). For (c)(d), $\mu_{R} = 4 \, \textrm{eV}$. For (a)(b)(d), $\Gamma = 0.4 \, \textrm{eV}$, $T = 300 \, \textrm{K}$.
}
\label{fig:AMR}
\end{figure}

Fig.~\ref{fig:AMR}(a) shows the intensity plot of total AM radiation as a function of the two chemical potentials $\mu_{L}$ and $\mu_{R}$ in the ortho position. We observe a four-line-segment feature (FLSF) where the AM radiation is large when one chemical potential is in resonance with the eigenstates at $\pm t$ and the bias window covers the energy range $[-t, t]$. The AM radiation is quite small in other regions. To analyze this resonant effect, we show in Fig.~\ref{fig:AMR}(b) the frequency/energy resolved spectrum of the AM radiation (Eq.~(\ref{eq:MonoDLDt}) without frequency integration) for the resonant case.  There is a large peak at $\hbar \omega = 2 t$, which implies that the FLSF in Fig.~\ref{fig:AMR}(a) is contributed mainly from the radiative transition processes $l = 2 \to l = 1$ and  $l = -2 \to l = -1$. 

To verify this further, we performed a simplified calculation using only four modes at $E = \pm t$. We find that the cross correlations between the degenerate states, such as $\widetilde{g}_{15}^{r}(E)$ and $\widetilde{g}_{24}^{r}(E)$, are important in resulting the net AM radiation than the correlations between nondegenerate states, such as $\widetilde{g}_{12}^{r}(E)$ and $\widetilde{g}_{14}^{r}(E)$. We set the latter to 0 for simplicity. With these simplifications, for the resonant peak at $\mu_{L} = -t$, we get from Eq.~(\ref{eq:MonoDLDt}) in the zero temperature limit~\cite{Supplementdetial}
\begin{equation} \label{eq:RET0}
\frac{d L_{z}}{d t} \approx J_{0} \frac{(\Gamma / 6)^2}{(\mu_{L} + t)^2 + (\Gamma / 6)^2} ,
\end{equation}
with $J_{0} = \frac{2}{\sqrt{3}\pi} t \alpha (v_0 / c)^2$, $v_{0} = a t / \hbar$, the fine structure constant $\alpha = e^2 / (4 \pi \varepsilon_{0} \hbar c)$, and the light speed $c$. The approximate expression of Eq.~(\ref{eq:RET0}) agrees well with numerical results from Eq.~(\ref{eq:MonoDLDt}) [Fig.~\ref{fig:AMR}(c)].    Eq.~(\ref{eq:RET0}) implies that the height of the resonance peak is a constant at zero temperature, and its width is characterized by $\Gamma / 6$. 

Fig.~\ref{fig:AMR}(d) shows the AM radiation for different ways of connecting the leads.  For ortho position, the resonant peak is broadened and its height is reduced, compared to the zero-temperature result (\romannumeral2)  in Fig.~\ref{fig:AMR}(c). This is due to a higher temperature at $300 \, \textrm{K}$.  For the asymmetric couplings at meta position, the AM radiation shows resonant effect, similar to the result at ortho position but with opposite direction of AM.   However, for the symmetric couplings at para position, the net AM is 0 despite of the applied biases. These properties imply the  possibility to design smart optical devices that can control the generation of AM radiation by simply applying an electric bias to the molecule junction, a convenient way compared with controlling AM radiation using a temperature bias~\cite{Maghrebi2019, Khandekar2019} or an external magnetic field~\cite{OttCircular2018}.  


{\emph{Conclusion.--}}
In summary, using the NEGF method, we have developed a theoretical framework to study AM radiation from current-carrying molecular junctions. As an application, the theory identifies from a quantum viewpoint that electrons of a ring-like benzene molecule emit light with AM due to radiative transitions between different angular momentum states. Due to asymmetric couplings to the leads and electron tunneling between degenerate energy states with opposite angular momentum, large resonant effect with the bias potential was discovered. Our theory can be straightforwardly applied to more realistic chiral molecules.

\begin{acknowledgments}
Z.Q.Z. and J.S.W. acknowledge the support by MOE tier 2 Grant No. R-144-000-411-112 and FRC Grant No. R-144-000-402-114. J.T.L. is supported by the National Natural Science Foundation of China (Grant No. 21873033).
\end{acknowledgments}

\bibliographystyle{apsrev4-1}
\bibliography{RadMolecule}

\widetext
\clearpage
\begin{center}
\textbf{\large Supplemental Material: Angular momentum radiation from current-carrying molecular junctions}
\end{center}
\begin{center}
Zu-Quan Zhang, Jing-Tao L\"u and Jian-Sheng Wang 
\end{center}
\setcounter{equation}{0}
\setcounter{figure}{0}
\setcounter{table}{0}
\setcounter{page}{1}
\makeatletter
\renewcommand{\theequation}{S\arabic{equation}}
\renewcommand{\thefigure}{S\arabic{figure}}
\renewcommand{\bibnumfmt}[1]{[S#1]}
\renewcommand{\citenumfont}[1]{S#1}
\section{Derivation of the radiation formulas}
In this section, we give some details of the derivations of Eq.~(\ref{eq:MonoP}) and Eq.~(\ref{eq:MonoDLDt}) for energy radiation and AM radiation respectively in the main text. The photon GF is defined as $D_{\mu \nu}(\bm{r},\bm{r}'; \tau, \tau') = - \frac{i}{\hbar} \langle T_{\tau} A_{\mu}(\bm{r}, \tau) A_{\nu}(\bm{r}', \tau') \rangle$, where the operators are in the Heisenberg representation, and $T_{\tau}$ is the time-order operator on the Keldysh contour.  The required lesser GF in Eqs.~(\ref{eq:Sperp})-(\ref{eq:BBGF}) in the main text is obtained by the Keldysh equation $D^{<} = D^{r} \Pi^{<} D^{a}$, with $D^{a} = (D^{r})^{\dag}$. Specifically, it is given by
\begin{equation} \label{eq:Ddxid}
D^{<}(\bm{r},\bm{r}', \omega) = d^{r}(\bm{r}, \bm{r}_{i}, \omega) \chi^{<}(\bm{r}_{i}, \bm{r}_{j}, \omega) d^{a}(\bm{r}_{j}, \bm{r}', \omega),
\end{equation}
with $\chi^{<}= \varepsilon^{-1}  \Pi^{<} (\varepsilon^{\dag})^{-1}$. Here, the matrix multiplication in coordinate component subscripts ($\mu, \nu = x, y, z$) and summation over sites $i, j$ are implied. We have defined the matrix function $\varepsilon =  \mathbb{1} - \Pi^{r} d^{r} $ and $\varepsilon^{\dag} = \mathbb{1} - d^{a} \Pi^{a}$, with $\mathbb{1}$ for the identity matrix. Excitations of the photons and screening effect are included in $\chi^{<}(\bm{r}_{i}, \bm{r}_{j}, \omega)$.
The retarded photon GF in free space is 
\begin{equation}  \label{eq:d0r}
d_{\mu \nu}^{r} (r, \omega) = - \frac{e^{q r}}{4 \pi \varepsilon_{0} r c^2} \big(\delta_{\mu \nu} - \frac{x_{\mu} x_{\nu}}{r^2} \big) - \frac{1}{4 \pi \varepsilon_{0} r c^2 }  \Big( \frac{e^{q r} - 1}{r^2 q^2} - \frac{e^{q r}}{ r q}  \Big) \big( \delta_{\mu \nu} - 3 \frac{x_{\mu} x_{\nu}}{r^2} \big), 
\end{equation}
where $c$ is the speed of light in free space, and $q = i \frac{\omega}{c}$. 

In calculating the energy radiation and AM radiation, the normal order is introduced to remove the contribution of zero-point motion.  The vector field operator is split into a positive frequency part and a negative frequency part, $A_{\mu} = A_{\mu}^{(+)} + A_{\mu}^{(-)}$. The rule of normal order is to put operators with positive frequency part on the right side of negative frequency part, i.e., $\big{\langle} \colon A_{\mu}^{(+)} A_{\mu}^{(-)} \colon \big{\rangle} =  \big{\langle} A_{\mu}^{(-)} A_{\mu}^{(+)} \big{\rangle} $.  We can write 
\begin{equation}
- \frac{i}{\hbar} \langle \colon  A_{\mu}(\bm{r} t) A_{\nu}(\bm{r}' t') \colon  \rangle= D_{\mu \nu}^{>}(\bm{r}, \bm{r'}; t, t') + D_{\nu \mu}^{(N)}(\bm{r}', \bm{r}; t' , t) - D_{\mu \nu}^{(A)}(\bm{r}, \bm{r'}; t, t').
\end{equation}
Here the normal order and anti-normal order correlation functions are written as
\begin{eqnarray}
D_{\mu \nu}^{(N)}(\bm{r}, \bm{r'}; t, t') = - \frac{i}{\hbar} \big{\langle} A_{\mu}^{(-)} (\bm{r} t) A_{\nu}(\bm{r}' t')^{(+)}  \big{\rangle},  \hspace{5mm}   D_{\mu \nu}^{(A)}(\bm{r}, \bm{r'}; t, t') = - \frac{i}{\hbar} \big{\langle} A_{\mu}^{(+)} (\bm{r} t) A_{\nu}(\bm{r}' t')^{(-)}  \big{\rangle}.
\end{eqnarray}
Applying $\bm{E}_{\bot} =  - \partial_{t} \bm{A}$ and $\bm{B} = \bm{\nabla} \times \bm{A}$ using the Coulomb gauge, and neglecting longitudinal electric field in the far-field region, we get 
\begin{equation} \label{eq:EBDGAN}
\langle \colon E_{\mu}(\bm{r} t) B_{\nu} (\bm{r}' t') \colon \rangle_{t' \to t} = \varepsilon_{\nu \delta \xi} \Big( - \frac{\partial}{\partial x_{\delta}'} \Big) \int_{-\infty}^{+\infty} \frac{d \omega}{2 \pi} \hbar \omega \Big[ D_{\mu \xi}^{>} (\bm{r}, \bm{r}', \omega) + D_{\xi \mu}^{(N)} (\bm{r}', \bm{r}, - \omega) - D_{\mu \xi}^{(A)} (\bm{r}, \bm{r}', \omega) \Big].
\end{equation}
Using the relations (see Ref.[S3])
\begin{equation} 
D_{\mu \nu}^{(N)}(\bm{r}, \bm{r'}, \omega) = \theta(-\omega) D_{\mu \nu}^{>}(\bm{r}, \bm{r'}, \omega), \hspace{5 mm} D_{\mu \nu}^{(A)}(\bm{r}, \bm{r'}, \omega) = \theta(\omega) D_{\mu \nu}^{>}(\bm{r}, \bm{r'}, \omega),
\end{equation}
we get from Eq.~(\ref{eq:EBDGAN}) 
\begin{equation}   \label{eq:EBDGG}
\langle \colon E_{\mu}(\bm{r} t) B_{\nu} (\bm{r}' t') \colon \rangle_{t' \to t} = \varepsilon_{\nu \delta \xi} \Big( - \frac{\partial}{\partial x_{\delta}'} \Big) \Bigg[ \int_{-\infty}^{0} \frac{d \omega}{2 \pi} \hbar \omega  D_{\mu \xi}^{>} (\bm{r}, \bm{r}', \omega) +  \int_{0}^{+\infty} \frac{d \omega}{2 \pi} \hbar \omega  D_{\xi \mu}^{>} (\bm{r}', \bm{r}, - \omega)  \Bigg].
\end{equation}
Applying the relations $D_{\mu \nu}^{>} (\bm{r}, \bm{r}', \omega) = D_{\nu \mu}^{<} (\bm{r}', \bm{r}, - \omega)$ and $D_{\mu \nu}^{<} (\bm{r}, \bm{r}', \omega) = - \big[ D_{\nu \mu}^{<} (\bm{r}', \bm{r}, \omega) \big]^{\ast}$, we write Eq.~(\ref{eq:EBDGG}) as 
\begin{equation}   \label{eq:EBDL}
\langle \colon E_{\mu}(\bm{r} t) B_{\nu} (\bm{r}' t') \colon \rangle_{t' \to t} = \varepsilon_{\nu \delta \xi}  \int_{0}^{\infty} \frac{d \omega}{2 \pi} \hbar \omega  2 \textrm{Re} \bigg[ - \frac{\partial}{\partial x_{\delta}'} D_{\mu \xi}^{<} (\bm{r}, \bm{r}', \omega) \bigg].
\end{equation}
Bringing Eq.~(\ref{eq:EBDL}) into Eq.~(\ref{eq:Poynting}), we get Eq.~(\ref{eq:Sperp}). Similarly, we can get Eqs.~(\ref{eq:EEGF}) and (\ref{eq:BBGF}) in the main text. 

We write the free space GF in Eq.~(\ref{eq:d0r}) and its real space partial derivative  to order $O(1/r^2)$ as follows
\begin{eqnarray}
d_{\mu \nu}^{r} (r, \omega) &=& C \frac{e^{qr}}{r} (\delta_{\mu \nu} - \hat{R} _{\mu} \hat{R}_{\nu}) - C \frac{e^{qr}}{r^2} \frac{1}{q} (\delta_{\mu \nu} - 3 \hat{R} _{\mu} \hat{R}_{\nu}) + O(1/r^3), \label{eq:drO2} \\
\frac{\partial}{\partial x_{\gamma}} d_{\mu \nu}^{r} (r, \omega) &=& C  \frac{e^{q r}}{r}  q \hat{R}_{\gamma} (\delta_{\mu \nu} - \hat{R} _{\mu} \hat{R}_{\nu}) - C  \frac{e^{q r}}{r^2}  \Big[ 2 \hat{R}_{\gamma}  (\delta_{\mu \nu} - 3 \hat{R} _{\mu} \hat{R}_{\nu})  + \delta_{\gamma \mu}  \hat{R}_{\nu} + \delta_{\gamma \nu} \hat{R}_{\mu} \Big] + O(1/r^3),  \label{eq:partdrO2}
\end{eqnarray}
with $C = -1/(4 \pi \varepsilon_0 c^2)$ and $\hat{R}_{\mu} = x_{\mu} / r$.  Noting the full GF in Eq.~(\ref{eq:Ddxid}), we calculate the partial derivative of the full GF in Eq.~(\ref{eq:Sperp}) as below
\begin{equation}   \label{eq:partD}
\frac{\partial}{\partial x_{\delta}'} D_{\nu \xi}^{<} (\bm{r}, \bm{r}' , \omega) =  d_{\nu \mu_1}^{r} (\bm{r}, \bm{r}_{i} , \omega) \chi_{\mu_{1} \mu_{2}}^{<} (\bm{r}_{i}, \bm{r}_{j} , \omega) \frac{\partial}{\partial x_{\delta}'} d_{\mu_{2} \xi}^{a} (\bm{r}_{j}, \bm{r}' , \omega),
\end{equation}
where Einstein summation rule is assumed. Using the relation $d^{a} = (d^{r})^{\dag}$, and making the monopole approximation 
$d_{\mu \nu}^{r} (\bm{r}, \bm{r}_{i}, \omega) = d_{\mu \nu}^{r} (\bm{r} - \bm{r}_{i}, \omega)   \approx d_{\mu \nu}^{r} (r, \omega)$, we write Eq.~(\ref{eq:partD}) as 
\begin{equation}   \label{eq:partDb}
\frac{\partial}{\partial x_{\delta}'} D_{\nu \xi}^{<} (\bm{r}, \bm{r}' , \omega) \approx  d_{\nu \mu_1}^{r} (r , \omega) \sum_{ij} \chi_{\mu_{1} \mu_{2}}^{<} (\bm{r}_{i}, \bm{r}_{j} , \omega) \frac{\partial}{\partial x_{\delta}'} \big[ d_{ \xi \mu_{2}}^{r} (r' , \omega) \big]^{\ast}.
\end{equation}
Let's first focus on the calculation of the energy radiation. We write the spherical  surface integration of energy radiation by Eq.~(\ref{eq:RadPower}) as solid angle integration 
\begin{equation} \label{eq:RadPangl}
P = \int d \Omega R^2 \bm{S} \bm{\cdot} \hat{\bm{R}},
\end{equation}
where $\Omega$ is the solid angle of the spherical surface. In the large distance limit, $R^2 \bm{S} \bm{\cdot} \hat{\bm{R}}$ in Eq.~(\ref{eq:RadPangl}) should be a quantity independent of the distance $R$ due to conservation of the radiation energy. Thus, calculation of the Poynting vector in Eq.~(\ref{eq:Sperp}) should be kept to order $O(1/r^2)$,  which implies that we only need to keep terms of order $O(1/r)$ in Eq.~(\ref{eq:drO2}) and Eq.~(\ref{eq:partdrO2}) for calculating the energy radiation. Eq.~(\ref{eq:partDb}) can thus be calculated as 
\begin{equation} \label{eq:partDO2}
\bigg[ \frac{\partial}{\partial x_{\delta}'} D_{\nu \xi}^{<} (\bm{r}, \bm{r}' , \omega) \bigg] \bigg{|}_{\bm{r} = \bm{r}'} \approx  C \frac{e^{qr}}{r} (\delta_{\nu \mu_{1}} - \hat{R} _{\nu} \hat{R}_{\mu_{1}}) \sum_{ij} \chi_{\mu_{1} \mu_{2}}^{<} (\bm{r}_{i}, \bm{r}_{j} , \omega) C  \frac{e^{ - q r}}{r}  (-q) \hat{R}_{\delta} (\delta_{\xi \mu_{2}} - \hat{R} _{\xi} \hat{R}_{\mu_{2}}).
\end{equation}
Using Eq.~(\ref{eq:partDO2}) to calculate Eq.~(\ref{eq:Sperp}) and bringing the result to Eq.~(\ref{eq:RadPangl}), we get the energy radiation power as
\begin{equation} \label{eq:rsP}
P = -  \int_{0}^{\infty} \frac{d \omega}{2 \pi} \frac{\hbar \omega^2 }{8 \pi^2 \varepsilon_{0} c^3}  \int d \Omega (\delta_{\mu \nu} - \hat{R}_{\mu} \hat{R}_{\nu})  \sum_{ij} \textrm{Im} \big[ \chi_{\mu \nu}^{<} (\bm{r}_{i}, \bm{r}_{j} , \omega) \big]. 
\end{equation}
We can make the approximation $\chi_{\mu \nu}^{<}  (\bm{r}_{i}, \bm{r}_{j} , \omega) \approx \Pi_{\mu \nu}^{<} (\bm{r}_{i}, \bm{r}_{j} , \omega) $ considering that the screening effect due to the radiation field is very small and can be neglected. Performing the solid angle integration in Eq.~(\ref{eq:rsP}), we get
\begin{equation}
P = - \int_{0}^{\infty} \frac{d \omega}{2 \pi}   \frac{\hbar \omega^2}{3 \pi \varepsilon_{0} c^3} \textrm{Im}\big[ \Pi_{\mu\mu} ^{\textrm{tot},<}(\omega) \big],
\end{equation}
with $\Pi_{\mu\nu} ^{\textrm{tot},<}(\omega) = \sum_{i j}  \Pi_{\mu \nu}^{<}(\bm{r}_{i},\bm{r}_{j}, \omega)$. This is Eq.~(\ref{eq:MonoP}) in the main text.

For the AM radiation, we write the surface integration in Eq.~(\ref{eq:AMFPower}) as the solid angle integration
\begin{equation}  \label{eq:dLdtOmg}
\frac{d \bm{L}}{dt} = \int d\Omega R^3 \langle : \hat{\bm{R}} \times \overleftrightarrow{{\bm{T}}} :\rangle \bm{\cdot} \hat{\bm{R}}.
\end{equation}
In the large distance limit, $R^3 \langle : \hat{\bm{R}} \times \overleftrightarrow{{\bm{T}}} :\rangle \bm{\cdot} \hat{\bm{R}}$ in Eq.~(\ref{eq:dLdtOmg}) should be independent of $R$ due to angular momentum conservation. This requires that the full GF and its partial derivative in Eq.~(\ref{eq:EEGF}) and Eq.~(\ref{eq:BBGF}) should be calculated in the order of $O(1/r^3)$. We need to keep the terms of order $O(1/r^2)$ as well as $O(1/r)$ in Eq.~(\ref{eq:drO2}) and Eq.~(\ref{eq:partdrO2}) to calculate Eq.~(\ref{eq:EEGF}) and Eq.~(\ref{eq:BBGF}).  The calculation is similar to that for the energy radiation. Bringing Eq.~(\ref{eq:EEGF}) and Eq.~(\ref{eq:BBGF}) to Eq.~(\ref{eq:dLdtOmg}), we get the AM radiation rate as
\begin{equation}  \label{eq:rsDLDt} 
\frac{d \bm{L}}{d t} =   \int_{0}^{\infty}  \frac{d \omega}{2 \pi}  \frac{\hbar \omega}{4 \pi^2 \varepsilon_{0} c^3} \int d \Omega \textrm{Re} \big[ \hat{\bm{R}} \times \bm{\Pi} ^{\textrm{tot},<}(\omega) \bm{\cdot} \hat{\bm{R}} \big]. 
\end{equation}
Performing the solid angle integration in Eq.~(\ref{eq:rsDLDt}), we obtain Eq.~(\ref{eq:MonoDLDt}) in the main text.

\section{The selection rules for the tight-binding benzene molecule}
In this section, we give some details of the derivation of the selection rules for the AM radiation of the TB benzene molecule in the main text. To get an intuitive physical picture from the viewpoint that a photon is emitted via a radiative transition by electrons from a state with higher energy to  a state with lower energy, we make the transformation from real space to mode space to analyze the radiation process. For the calculation of Eq.~(\ref{eq:MonoP}) and Eq.~(\ref{eq:MonoDLDt}) in the main text, the transformation to mode space is implemented as
\begin{equation}  \label{eq:Ximode}
\begin{split}
\Pi_{\mu\nu} ^{\textrm{tot},<}(\omega) =& - i \hbar   \int_{-\infty}^{+\infty} \frac{dE}{2\pi \hbar} \textrm{Tr}\Big[ M^{\mu} g^{<} (E) M^{\nu} g^{>} (E - \hbar \omega) \Big]  \\ 
=& - i \hbar   \int_{-\infty}^{+\infty} \frac{dE}{2\pi \hbar} \textrm{Tr}\Big[ U^{\dag} M^{\mu} U U^{\dag} g^{<} (E) U U^{\dag} M^{\nu} U U^{\dag}  g^{>} (E - \hbar \omega) U \Big]  \\
=& - i \hbar   \int_{-\infty}^{+\infty} \frac{dE}{2\pi \hbar} \textrm{Tr}\Big[ \widetilde{M}^{\mu} \widetilde{g}^{<}(E) \widetilde{M}^{\nu} \widetilde{g}^{>}(E - \hbar \omega) \Big].  
\end{split}
\end{equation}
Here $M^{\mu}$ is the electron-photon coupling matrix summed over the photon site index $k$. For the simplest case that only two eigenstates with energy $E_m$ and $E_n$ of the TB benzene molecule  are connected to the two leads respectively (see Fig.~\ref{fig:SRDOS}(b) in the main text).    
Eq.~(\ref{eq:Ximode}) can be written as
\begin{equation} \label{eq:XiMDb}
\begin{split}
\Pi_{\mu\nu} ^{\textrm{tot},<}(\omega) =& - i \hbar   \int_{-\infty}^{+\infty} \frac{dE}{2\pi \hbar}  \widetilde{M}_{mn}^{\mu} \widetilde{g}_{nn}^{<}(E) \widetilde{M}_{nm}^{\nu} \widetilde{g}_{mm}^{>}(E - \hbar \omega)   \\
 & - i \hbar   \int_{-\infty}^{+\infty} \frac{dE}{2\pi \hbar}  \widetilde{M}_{nm}^{\mu} \widetilde{g}_{mm}^{<}(E) \widetilde{M}_{mn}^{\nu} \widetilde{g}_{nn}^{>}(E - \hbar \omega).  
\end{split}
\end{equation}
In the high bias regime $\mu_{L} \ll E_{n} < E_{m} \ll \mu_{R}$, we make the approximations $\widetilde{g}_{mm}^{<}(E) \approx \frac{i \bar{\Gamma}}{(E - E_m)^2 + (\bar{\Gamma}/2)^2}$, $\widetilde{g}_{nn}^{>}(E - \hbar \omega) \approx \frac{- i \bar{\Gamma}}{(E - E_{n} - \hbar \omega)^2 + (\bar{\Gamma}/2)^2}$, $\widetilde{g}_{mm}^{>}(E - \hbar \omega) \approx 0$ and $\widetilde{g}_{nn}^{<}(E) \approx 0$. 
The first term in Eq.~(\ref{eq:XiMDb}) is zero, so we can write
 \begin{equation} \label{eq:XiMDc}
\Pi_{\mu\mu}^{\textrm{tot},<}(\omega) \approx - i \hbar   \int_{-\infty}^{+\infty} \frac{dE}{2\pi \hbar}  \widetilde{M}_{nm}^{\mu}  \widetilde{M}_{mn}^{\nu}  \widetilde{g}_{mm}^{<}(E) \widetilde{g}_{nn}^{>}(E - \hbar \omega).    
\end{equation}
Bringing Eq.~(\ref{eq:XiMDc}) into Eq.~(\ref{eq:MonoP}) and Eq.~(\ref{eq:MonoDLDt}), we  get
\begin{eqnarray}
P & = & \int_{0}^{\infty} \frac{d \omega}{2 \pi}  \frac{\hbar^2 \omega^2}{3 \pi \varepsilon_{0} c^3} \int_{-\infty}^{\infty} \frac{dE}{2\pi \hbar} \Big( \widetilde{M}_{nm}^{x}  \widetilde{M}_{mn}^{x} + \widetilde{M}_{nm}^{y}  \widetilde{M}_{mn}^{y}  \Big) \widetilde{g}_{mm}^{<}(E) \widetilde{g}_{nn}^{>}(E - \hbar \omega) , \label{eq:PA} \\
\frac{d L_{z}}{d t} & =&  \int_{0}^{\infty} \frac{d \omega}{2 \pi}  \frac{\hbar^2 \omega}{3 \pi \varepsilon_{0} c^3} \int_{-\infty}^{\infty} \frac{dE}{2\pi \hbar} \Big( -i \widetilde{M}_{nm}^{x}  \widetilde{M}_{mn}^{y} + i \widetilde{M}_{nm}^{y}  \widetilde{M}_{mn}^{x}  \Big) \widetilde{g}_{mm}^{<}(E) \widetilde{g}_{nn}^{>}(E - \hbar \omega).  \label{eq:LA}   
\end{eqnarray}
 
The benzene molecule model has $C_{6}$ symmetry. We choose the positions of the six sites as $x_{j} = a \, \textrm{cos}(2\pi j / 6)$, $y_{j} = a \, \textrm{sin}(2\pi j / 6)$, $z_{j} = 0$, with $j = 1, 2, \dots, 6$.  Noting $M_{ij}^{\mu} = i e v_{ij}^{\mu} = i \frac{e}{\hbar} t_{ij} (\bm{r}_{i} - \bm{r}_{j})_{\mu}$,  $U_{j m} = 1/ \sqrt{6} e^{i 2 \pi j m / 6}$, we get
\begin{equation}   \label{eq:MM}
\widetilde{M}^{x} = \frac{e a t}{2 \hbar} \left[ {\begin{array}{cccccc}
0   & 2 i  & 0  & 0  & 0   & - i  \\
-2i &   0  & i   & 0  & 0   & 0 \\
0   &  -i   & 0  & -i  & 0   & 0 \\
0   &  0   & i   & 0  & - 2i &  0 \\
0   &  0   & 0   & 2i  & 0 &  -i \\
i   &  0   & 0   & 0  & i &  0  \\
\end{array} }  \right],  \hspace{5mm} 
\widetilde{M}^{y} = \frac{e a t}{2 \hbar} \left[ {\begin{array}{cccccc}
0   & -2   & 0  & 0  & 0   & -1  \\
-2 &   0  & -1   & 0  & 0   & 0 \\
0   &  -1  & 0  & 1  & 0   & 0 \\
0   &  0   & 1   & 0  & 2 &  0 \\
0   &  0   & 0   & 2  & 0 &  1 \\
-1   &  0   & 0   & 0  & 1 &  0  \\
\end{array} }  \right],
\end{equation}

\begin{equation}   \label{eq:MtM}
(\widetilde{M}^{x})^{T}  \bm{\cdot}  \widetilde{M}^{x}  = \Big( \frac{e a t}{2 \hbar} \Big)^2 \left[ {\begin{array}{cccccc}
0   & 4  & 0  & 0  & 0   & 1  \\
4 &   0  & 1   & 0  & 0   & 0 \\
0   &  1  & 0  & 1  & 0   & 0 \\
0   &  0   & 1   & 0  & 4 &  0 \\
0   &  0   & 0   & 4 & 0 &  1 \\
1   &  0   & 0   & 0  & 1 &  0  \\
\end{array} }  \right],  \hspace{5mm} 
(\widetilde{M}^{x})^{T}  \bm{\cdot}  \widetilde{M}^{y}  = \Big( \frac{e a t}{2 \hbar} \Big)^2 \left[ {\begin{array}{cccccc}
0   & 4 i  & 0  & 0  & 0   & - i  \\
- 4 i &   0  & i   & 0  & 0   & 0 \\
0   &  - i  & 0  & i  & 0   & 0 \\
0   &  0   & - i   & 0  &  4 i &  0 \\
0   &  0   & 0   & - 4 i & 0 &  i \\
i   &  0   & 0   & 0  & -i &  0  \\
\end{array} }  \right].
\end{equation}
Here, we have defined an element-wise matrix multiplication $\Big [ (\widetilde{M}^{\mu})^{T}  \bm{\cdot}  \widetilde{M}^{\nu}  \Big]_{mn} =  \widetilde{M}_{nm}^{\mu} \widetilde{M}_{mn}^{\nu}$. From Eq.~(\ref{eq:MM}) and Eq.~(\ref{eq:MtM}), we get the relations 
\begin{equation}  \label{eq:SmtM}
\widetilde{M}_{nm}^{x} \widetilde{M}_{mn}^{x} = \widetilde{M}_{nm}^{y} \widetilde{M}_{mn}^{y},    \hspace{5mm}   \widetilde{M}_{nm}^{x} \widetilde{M}_{mn}^{y} = i \Delta_{mn} \widetilde{M}_{nm}^{x} \widetilde{M}_{mn}^{x}. 
\end{equation}
Here, $\Delta_{mn} = \textrm{sgn}(m - n)$ if $|m - n| = 1$, and $\Delta_{16} = - \Delta_{61} = 1$, otherwise $\Delta_{mn} = 0$. Taking the weak coupling limit $\bar{\Gamma} \to 0 $ in Eq.~(\ref{eq:PA}) and Eq.~(\ref{eq:LA}), we get
\begin{eqnarray}
P & = & \frac{\omega_{mn}^2}{3 \pi \varepsilon_{0} c^3} \Big( \widetilde{M}_{nm}^{x}  \widetilde{M}_{mn}^{x} + \widetilde{M}_{nm}^{y}  \widetilde{M}_{mn}^{y}  \Big) , \label{eq:PAGM0} \\
\frac{d L_{z}}{d t} & =&  \frac{\omega_{mn}}{3 \pi \varepsilon_{0} c^3} \Big( -i \widetilde{M}_{nm}^{x}  \widetilde{M}_{mn}^{y} + i \widetilde{M}_{nm}^{y}  \widetilde{M}_{mn}^{x}  \Big),  \label{eq:LAGM0}   
\end{eqnarray}
with $\hbar \omega_{mn} = E_{m} - E_{n}$. Bringing Eq.~(\ref{eq:SmtM}) into Eq.~(\ref{eq:PAGM0}) and Eq.~(\ref{eq:LAGM0}), we get 
\begin{equation} \label{eq:LEratio}
\frac{d L_{z} / d t}{P} = \frac{\Delta_{mn}}{\omega_{mn}}.
\end{equation}
Since every emitted photon carries an energy $\hbar \omega_{mn}$, the number of photons emitted per unit time is $d N / d t = P / (\hbar \omega_{mn})$. 
The angular momentum carried by an emitted photon is $\Delta L = \frac{dL_z / dt}{dN / d t} = \Delta_{mn} \hbar$.

\section{Derivation of the resonant effect}
In this section, we give a derivation of Eq.~(\ref{eq:RET0}) in the main text for the resonant effect of the AM radiation. The resonant effect is supposed to involve the inelastic transitions $l = 2 \to l = 1$ and $l = -2 \to l = -1$ with the emitted photon energy $\hbar \omega \approx 2 t$, supported by the FLSF with the chemical potential bias $|\mu_{L} - \mu_{R}| = 2 t$ [Fig.~\ref{fig:AMR}(a)] and the AM radiation spectrum with the large peak at $\hbar \omega = 2 t$ [Fig.~\ref{fig:AMR}(b)]. The interacting self-energy including these two processes is written as 
\begin{equation}
\begin{split}
\Pi_{\mu \nu}^{\textrm{tot},<}(\omega) =&  - i \hbar   \int_{-\infty}^{+\infty} \frac{dE}{2\pi \hbar}  \widetilde{M}_{12}^{\mu} \widetilde{g}_{22}^{<}(E) \widetilde{M}_{21}^{\nu} \widetilde{g}_{11}^{>}(E - \hbar \omega) \\
&  - i \hbar   \int_{-\infty}^{+\infty} \frac{dE}{2\pi \hbar}  \widetilde{M}_{54}^{\mu} \widetilde{g}_{44}^{<}(E) \widetilde{M}_{45}^{\nu} \widetilde{g}_{55}^{>}(E - \hbar \omega). 
\end{split}
\end{equation}
Noting the $\widetilde{M}^{\mu}$ matrix in Eq.~(\ref{eq:MM}),  we get 
\begin{equation} \label{eq:PIxy}
\begin{split}
\Pi_{x y}^{\textrm{tot},<}(\omega) - \Pi_{y x}^{\textrm{tot},<}(\omega) =& - 2 e^2 v_{0}^2  \int_{-\infty}^{+\infty} \frac{dE}{2\pi} \Big[ \widetilde{g}_{22}^{<}(E) \widetilde{g}_{11}^{>}(E - \hbar \omega)  - \widetilde{g}_{44}^{<}(E) \widetilde{g}_{55}^{>}(E - \hbar \omega) \Big],
\end{split}
\end{equation}
with $v_{0} = a t / \hbar$. We consider only the cross correlations between the degenerate states, which is important to account for the resonant effect. Specifically, for degenerate states $l = 1$ and $l = -1$, we write the GFs as
\begin{equation}  \label{eq:GF15}
\left[ {\begin{array}{cc}
\widetilde{g}_{11}^{r}(E)   & \widetilde{g}_{15}^{r} (E)  \\
\widetilde{g}_{51}^{r}(E)   & \widetilde{g}_{55}^{r} (E)  \\
\end{array} }  \right]  = \left[ {\begin{array}{cc}
E +  t  - \big[ \widetilde{\Sigma}_{\textrm{leads}}^{r}\big]_{11}  & - \big[\widetilde{\Sigma}_{\textrm{leads}}^{r}\big]_{15}   \\
- \big[ \widetilde{\Sigma}_{\textrm{leads}}^{r}\big]_{51}   &E + t  - \big[ \widetilde{\Sigma}_{\textrm{leads}}^{r}\big]_{55}   \\
\end{array} }  \right]^{-1}, 
\end{equation}
and
\begin{equation}
\left[ {\begin{array}{cc}
\widetilde{g}_{11}^{<}   & \widetilde{g}_{15}^{<}   \\
\widetilde{g}_{51}^{<}   & \widetilde{g}_{55}^{<}   \\
\end{array} }  \right]  = \left[ {\begin{array}{cc}
\widetilde{g}_{11}^{r}   & \widetilde{g}_{15}^{r}   \\
\widetilde{g}_{51}^{r}   & \widetilde{g}_{55}^{r}   \\
\end{array} }  \right] 
\left[ {\begin{array}{cc}
\big[\widetilde{\Sigma}_{\textrm{leads}}^{<}\big]_{11}   & \big[\widetilde{\Sigma}_{\textrm{leads}}^{<}\big]_{15}   \\
\big[\widetilde{\Sigma}_{\textrm{leads}}^{<}\big]_{51}   & \big[\widetilde{\Sigma}_{\textrm{leads}}^{<}\big]_{55}   \\
\end{array} }  \right] 
 \left[ {\begin{array}{cc}
\widetilde{g}_{11}^{a}   & \widetilde{g}_{15}^{a}   \\
\widetilde{g}_{51}^{a}   & \widetilde{g}_{55}^{a}   \\
\end{array} }  \right]. 
\end{equation}

We consider the leads are coupled to the benzene molecule in the ortho position. The retarded lead self-energy is $ \big[\Sigma_{\textrm{leads}}^{r}\big]_{ij} = - i \frac{\Gamma}{2} \delta_{ij} (\delta_{i1} + \delta_{i2})$. The lesser and greater self-energies are $\big[ \Sigma_{\textrm{leads}}^{<} \big]_{11} = i f_{\textrm{L}} \Gamma$,  $\big[ \Sigma_{\textrm{leads}}^{<} \big]_{22} = i f_{\textrm{R}} \Gamma$, $\big[ \Sigma_{\textrm{leads}}^{>} \big]_{11} = i (-1 + f_{L}) \Gamma$, $\big[ \Sigma_{\textrm{leads}}^{>} \big]_{22} = i (-1 + f_{R}) \Gamma$. The mode space self-energies can be obtained by $\widetilde{\Sigma}_{\textrm{leads}}^{r,<, >} = U^{\dag} \Sigma_{\textrm{leads}}^{r,<, >} U $. We get from Eq.~(\ref{eq:GF15})
\begin{eqnarray}
\widetilde{g}_{11}^{r}(E) &=& \frac{1}{2} \Big[ \frac{1}{E+t + i \Gamma / 12} +  \frac{1}{E+t + i \Gamma / 4}  \Big]  \approx  \frac{1}{E+t + i \Gamma / 6},  \\
\widetilde{g}_{15}^{r}(E) &=& \frac{1}{2} \Big[ \frac{1}{E+t + i \Gamma / 12} -  \frac{1}{E+t + i \Gamma / 4}  \Big]  \approx  \frac{i \Gamma / 12}{(E+t + i \Gamma / 6)^2},
\end{eqnarray}
and $\widetilde{g}_{55}^{r}(E) = \widetilde{g}_{11}^{r}(E)$.
Similarly, we get 
\begin{eqnarray}
\widetilde{g}_{22}^{r}(E) &\approx&  \frac{1}{E - t + i \Gamma / 6},  \\
\widetilde{g}_{24}^{r}(E) &\approx& \frac{i \Gamma / 12}{(E - t + i \Gamma / 6)^2},
\end{eqnarray}
and $\widetilde{g}_{44}^{r}(E) = \widetilde{g}_{22}^{r}(E)$. We obtain
\begin{equation} \label{eq:G2145}
\begin{split}
&  \widetilde{g}_{22}^{<}(E) \widetilde{g}_{11}^{>}(E - \hbar \omega)  - \widetilde{g}_{44}^{<}(E) \widetilde{g}_{55}^{>}(E - \hbar \omega) \\
\approx & \frac{\Gamma^2}{6 \sqrt{3}} \textrm{Im} \Big[ \big( \widetilde{g}_{24}^{r}(E) \big)^{\ast} \widetilde{g}_{22}^{r}(E)  \Big] \big| \widetilde{g}_{11}^{r}(E - \hbar \omega)  \big|^2 \Big[ f_{L}(E) - f_{R}(E)  \Big] \Big[ -2 + f_{L}(E- \hbar \omega) + f_{R}(E - \hbar \omega) \Big]  \\
& - \frac{\Gamma^2}{6 \sqrt{3}} \textrm{Im} \Big[ \big( \widetilde{g}_{15}^{r}(E - \hbar \omega) \big)^{\ast} \widetilde{g}_{11}^{r}(E - \hbar \omega)  \Big] \big| \widetilde{g}_{22}^{r}(E)  \big|^2 \Big[ f_{L}(E - \hbar \omega) - f_{R}(E - \hbar \omega)  \Big] \Big[f_{L}(E) + f_{R}(E) \Big].
\end{split}
\end{equation}
In getting Eq.~(\ref{eq:G2145}), we have omitted some higher order terms considering that we may take the cross correlations of degenerate states $\widetilde{g}_{15}^{r}$ and $\widetilde{g}_{24}^{r}$ as perturbations. Calculating Eq.~(\ref{eq:PIxy}) using Eq.~(\ref{eq:G2145}) and bringing the result to Eq.~(\ref{eq:MonoDLDt}),  we get the AM radiation
\begin{equation} \label{eq:Resonant4}
\begin{split}
\frac{d L_{z}}{d t} \approx & J_{0} \Bigg[ \theta(-t - \mu_{R}) \frac{ (\Gamma / 6)^2 }{ (\mu_{L} - t)^2 + (\Gamma / 6)^2 } -  \theta(-t - \mu_{L}) \frac{ (\Gamma / 6)^2 }{ (\mu_{R} - t)^2 + (\Gamma / 6)^2 } \\
& + \theta(\mu_{R} - t) \frac{ (\Gamma / 6)^2 }{ (\mu_{L} + t)^2 + (\Gamma / 6)^2 } -  \theta(\mu_{L} - t) \frac{ (\Gamma / 6)^2 }{ (\mu_{R} + t)^2 + (\Gamma / 6)^2 } \Bigg],
\end{split}
\end{equation}
with $J_{0} = \frac{2}{\sqrt{3}\pi} t \alpha (v_0 / c)^2$ and $\alpha = e^2 / (4 \pi \varepsilon_{0} \hbar c)$. In getting Eq.~(\ref{eq:Resonant4}), we have used zero temperature limit for the Fermi functions of the two leads and assumed the lead coupling is weak, i.e., $\Gamma \ll t$. Taking $\mu_{R} = 4 \, \textrm{eV} > t$ in Eq.~(\ref{eq:Resonant4}), we get Eq.~(\ref{eq:RET0}) in the main text.

\vskip 12pt

\noindent [S1] O. Keller, {\sl Quantum theory of near-field electrodynamics}, (Springer, Berlin, Germany, 2012). \\
\noindent [S2] K. Kaasbjerg and A. Nitzan, Phys. Rev. Lett. {\bf 114}, 126803 (2015). \\
\noindent [S3] M. Janowicz, D. Redding, and M. Holthaus, Phys. Rev. A {\bf 68}, 043823 (2003). \\
\noindent [S4] M. Katoh, M. Fujimoto, H. Kawaguchi, K. Tsuchiya, K. Ohmi, T. Kaneyasu, Y. Taira, M. Hosaka, A. Mochihashi, and Y. Takashima, Phys. Rev. Lett. {\bf 118}, 094801 (2017). 

\end{document}